\documentclass[12pt]{iopart}
\usepackage[colorlinks=true,urlcolor=blue,linkcolor=blue]{hyperref}
\usepackage{graphicx
}
\relax

\newcommand{\bea}{\begin{eqnarray}}
\newcommand{\bdm}{\begin{displaymath}}
\newcommand{\edm}{\end{displaymath}}
\newcommand{\nn}{\nonumber}
\newcommand{\eea}{\end{eqnarray}}

\def\be{\begin{equation}}
\def\ee{\end{equation}}
\def\bs{\begin{subequations}}
\def\es{\end{subequations}}
\def\calm{{\cal M}}
\def\calmb{{\bar{\cal M}}}

\def\lx{\lambda}
\def\ex{\epsilon}

\newcommand{\sx}{\sigma}

\newcommand{\Ab}{{\bar A}}
\newcommand{\Omb}{{\bar \Omega}}
\newcommand{\Hb}{{\bar H}}
\newcommand{\rb}{{\bar r}}

\newcommand{\lb}{{\bar \lx}}
\newcommand{\sxb}{{\bar \sx}}
\newcommand{\thb}{{\bar \theta}}
\newcommand{\tb}{{\bar \tau}}
\newcommand{\ttb}{{\bar t}}
\newcommand{\fb}{{\bar f}}
\newcommand{\Rb}{{\bar R}}

\newcommand{\rhb}{\bar{\rho}}

\newcommand{\kb}{\bar{k}}

\newcommand{\dx}{{\rm d}}

\def\be{\begin{equation}}
\def\ee{\end{equation}}
\def\bs{\begin{subequations}}
\def\es{\end{subequations}}
\newcommand{\een}{\end{subequations}}
\newcommand{\ben}{\begin{subequations}}
\newcommand{\beq}{\begin{eqalignno}}
\newcommand{\eeq}{\end{eqalignno}}

\def \lta {\mathrel{\vcenter
     {\hbox{$<$}\nointerlineskip\hbox{$\sim$}}}}
\def \gta {\mathrel{\vcenter
     {\hbox{$>$}\nointerlineskip\hbox{$\sim$}}}}

\newcommand\fverb{\setbox\pippobox=\hbox\bgroup\verb}
\newcommand\fverbdo{\egroup\medskip\noindent%
                        \fbox{\unhbox\pippobox}\ }
\newcommand\fverbit{\egroup\item[\fbox{\unhbox\pippobox}]}
\newbox\pippobox

\def \lta {\mathrel{\vcenter
     {\hbox{$<$}\nointerlineskip\hbox{$\sim$}}}}
\def \gta {\mathrel{\vcenter
     {\hbox{$>$}\nointerlineskip\hbox{$\sim$}}}}

\begin{document}

\title{The Effect of Large-Scale Inhomogeneities on the Luminosity Distance}

\author{Nikolaos Brouzakis, Nikolaos Tetradis and Eleftheria Tzavara}
\address{
University of Athens, Department of Physics, University Campus, 
Zographou 157 84, Athens, Greece}

\begin{abstract}
We study the form of the luminosity distance as a function of redshift
in the presence of large scale inhomogeneities, with
sizes of order 10 Mpc or larger. We approximate the Universe
through the Swiss-cheese model, with each spherical region described by
the Lemaitre-Tolman-Bondi metric. 
We study the propagation of light beams in this
background, assuming that the locations of the 
source and the observer are random.
We derive the optical equations for the evolution of the
beam area and shear.  
Through their integration we determine the configurations that can lead to 
an increase of the luminosity distance relative to the homogeneous cosmology.
We find that this can be achieved if the Universe is composed of 
spherical void-like regions, with matter concentrated near their
surface. For inhomogeneities consistent with the observed 
large scale structure,
the relative increase of the luminosity distance is of the order 
of a few percent at redshifts near 1, 
and falls short of explaining the substantial increase 
required by the supernova data. On the other hand, the effect we 
describe is important for the correct determination of the energy content of
the Universe from observations.
\end{abstract}
\maketitle

\section{Introduction}
\setcounter{equation}{0}

The form of the luminosity distance as a function of redshift for
distant supernovae supports the conclusion that the recent expansion of 
our Universe has been accelerating
\cite{accel1,accel2}. 
In the context of homogeneous cosmology, 
a recent accelerating phase is also
in agreement with the observed perturbations in the cosmic microwave
background \cite{wmap}.
The simplest explanation for this phenomenon is that the
cosmological constant is non-zero. However, the absence of acceleration at
redshifts $z\gta 1$ implies that the required value of the cosmological 
constant is approximately 120 orders of magnitude smaller than its natural
value in terms of the Planck scale. 

Various attempts at an alternative explanation try to build a link 
between inhomogeneities in the matter distribution in the Universe 
and the perceived cosmological acceleration. 
For example, there have been
arguments, based on perturbative estimates, that the 
backreaction of super-horizon
perturbations on the cosmological expansion 
is significant and could cause the acceleration \cite{kolb}. 
However, the validity of this 
effect is questionable \cite{seljak,rasanen1}.

We are interested in the importance
for the problem of cosmological acceleration
of inhomogeneities with sub-horizon characteristic scales today 
\cite{rasanenn}.
An intriguing fact is that 
the accelerating phase coincides with the 
period in which inhomogeneities in the matter distribution at scales 
${\cal O} (10)\, h^{-1}$ Mpc 
become large, so that the Universe cannot be approximated
as homogeneous any more at these scales. Such distances are
still small compared to the Hubble distance 
$\sim 3\times 10^3\,h^{-1}$ Mpc, that
sets the scale over which the Universe is assumed to be described by
the homogeneous Friedmann-Robertson-Walker (FRW) metric.
However, the presence of structures, such as walls or voids, 
with sizes up to $100\,h^{-1}$ Mpc poses the question of whether their 
influence on the homogeneous solution is substantial.

Because of the significant 
growth of such inhomogeneities at recent times, a perturbative
treatment may not be sufficient. An exact solution of the Einstein 
equations, even for a simplified geometry, could be more useful in order to
reveal an underlying mechanism. The 
Lemaitre-Tolman-Bondi (LTB) metric \cite{ltb} has
been employed often in this context 
\cite{mustapha}--\cite{biswas}. It has been observed that 
any form of the luminosity distance as a function of redshift can be 
reproduced with this metric \cite{mustapha}. However, reproducing the 
supernova data requires a variation of the 
density or the expansion rate over distances 
${\cal O}(10^3)\,h^{-1}$ Mpc \cite{alnes}--\cite{biswas}.
Moreover, in order to avoid a conflict with the isotropy of the 
Cosmic Microwave Background (CMB), the location
of the observer must be near the center of the spherical configuration 
described by the LTB metric.

In this work we study the effect of inhomogeneities on the luminosity 
distance without assuming a preferred location of the observer. 
We model the inhomogeneities as spherical regions within which the geometry
is described by the LTB metric. At the boundary of
these regions, the LTB metric is matched with the FRW metric
that describes the evolution in the region between the inhomogeneities.
In this way, we model the Universe as consisting of collapsing or
expanding inhomogeneous regions, while we preserve the notion of 
a common scale factor that describes the expansion of the homogeneous
intermediate regions. Our model is similar to the standard Swiss-cheese
model \cite{swiss}, with the replacement of the Schwarzschild metric with 
the LTB one. For this reason we refer to it as the LTB Swiss-cheese model.

The transmission of a light beam in such a background can be studied
through the Sachs optical equations \cite{sachs}. These describe the expansion
and shear of the beam along its null trajectory. Apart from the case of
an FRW background, the optical equations have been derived by Kantowski 
for a Scharzschild background in his study of light 
propagation in the Swiss-cheese model \cite{kantowski}.
We derive the equations for a general LTB background. We then use them in order
to study light propagation in the LTB Swiss-cheese model. We focus on the
modification of the luminosity distance as a function of redshift with 
respect to the homogeneous FRW case. We investigate which configurations
lead to an increase of the luminosity distance, and estimate the magnitude
of the effect.

In the following section and in the appendices we derive the optical
equations for LTB, FRW and Schwarzschild backgrounds. The expressions for
the last two cases are known, but we summarize them for completeness.
In section 3 we discuss the particular details of the LTB metric that we
use for the description of the spherical regions of the LTB Swiss-cheese model.
In section 4 we present numerical solutions for the cosmological evolution of
the background metric. In section 5 we discuss light propagation and the
modification of the luminosity distance in the inhomogeneous background. 
In section 6 we estimate the magnitude of this modification. In section
7 we present our conclusions on the possible explanation of the form of the
luminosity function in the context of inhomogeneous cosmology.

\section{Optical equations and luminosity distance}
\setcounter{equation}{0}

The proper cross-section area A of a light beam obeys the equation
\be
\frac{dA}{d\lambda}=2\theta A.
\label{tha} \ee
where $\theta$ is the expansion of the beam and $\lx$ an affine parameter
along the null beam trajectory.
The symmetric and traceless shear tensor
\be
\sigma_{ab}=\left( \begin{array}{cc}
\sx_1 & \sigma_2 \\
\sigma_2 & -\sx_1 \\
\end{array} \right)
\label{sxab}
\ee
describes deformations of the beam. 
In this work we study light propagation in
space-times with spherical spatial symmetry. In such cases, the
off-diagonal elements $\sx_2$ of the shear tensor can be set consistently
to zero. The reason is that the eigenvalues $\pm \sx$ of the 
shear tensor determine the deformation of a surface along two principal
orthogonal
axes perpendicular to the light direction. The rate of stretching 
in these two directions is given by $\theta+\sx$ and $\theta-\sx$, 
respectively. 
In the case of spherical symmetry, one principal
axis always lies on the plane determined by the null geodesic and the center of
symmetry and the second perpendicularly to it. 
It is possible then to choose 
a reference basis that includes unit vectors along these principal
directions. This permits us to 
set $\sx_2=0$, $\sx=\sx_1$.

The optical equations, derived by Sachs \cite{sachs}, 
determine the evolution of the beam expansion and shear as the light
propagates in a background geometry. For completeness, we include 
a derivation of these equations in appendix A. In appendix B we derive their
specific form in various backgrounds that are relevant for our study. 

\subsection{Lemaitre-Tolman-Bondi (LTB) background}

Under the assumption of spherical symmetry,
the most general metric for a 
pressureless, inhomogeneous fluid is the  
LTB metric \cite{ltb}.
It can be written in the form 
\begin{equation}
ds^{2}=-dt^2+b^2(t,r)dr^2+R^2(t,r)d\Omega^2,
\label{metrictb}
\end{equation}
where $d\Omega^2$ is the metric on a two-sphere. 
The function $b(t,r)$ is given by
\begin{equation}
b^2(t,r)=\frac{R'^2(t,r)}{1+f(r)},
\label{brttb} \end{equation}
where the prime denotes differentiation with respect to $r$, and
$f(r)$ is an arbitrary function.
The bulk energy momentum tensor has the form
\begin{equation}
T^A_{~B}={\rm diag} \left(-\rho(t,r),\, 0,\, 0,\, 0  \right).
\label{enmomtb} \end{equation}
The fluid consists of successive shells marked by $r$, whose
local density $\rho$ is time-dependent. 
The function $R(t,r)$ describes the location of the shell marked by $r$
at the time $t$. Through an appropriate rescaling it can be chosen to satisfy
\be
R(0,r)=r.
\label{s0} \ee

The Einstein equations reduce to 
\begin{eqnarray}
\dot{R}^2(t,r)&=&\frac{1}{8\pi M^2}\frac{\calm (r)}{R}+f(r)
\label{tb1} \\
\calm'(r)&=&4\pi R^2 \rho \, R',
\label{tb2} \end{eqnarray}
where the dot denotes differentiation with respect to $t$, and 
$G=\left( 16 \pi M^2 \right)^{-1}$.
The generalized mass function $\calm(r)$ of the fluid can be chosen 
arbitrarily. It incorporates the
contributions of all shells up to $r$. It determines the energy density
through eq. (\ref{tb2}). Because of energy conservation
$\calm(r)$  
is independent of $t$, while $\rho$ and $R$ depend on both $t$ and $r$.

Without loss of generality we consider 
geodesic null curves on the plane with $\theta=\pi/2$. 
The geodesic equations are ($k^i={dx^i}/{d\lambda}$)
\bea
&&\frac{dk^0}{d\lx}+\frac{\dot{R}'R'}{1+f}\left(k^1\right)^2+
\dot{R}R\left( k^3\right)^2=0 \label{gtb1}\\
&&\frac{dk^1}{d\lx}+2\frac{\dot{R}'}{R'}k^0
k^1+\left(\frac{R''}{R'}-\frac{f'}{2(1+f)}\right)
\left( k^1\right)^2-(1+f)\frac{R}{R'}\left(k^3\right)^2=0\label{gtb2}\\
&&\frac{dk^3}{d\lx}+2\frac{\dot{R}}{R}k^0
k^3+2\frac{R'}{R}k^1 k^3=0.
\label{gtb3}\eea
One of them can be replaced by the null condition
\be
-\left(k^0\right)^2+\frac{R'^2}{1+f}\left(k^1\right)^2+R^2
\left( k^3\right)^2=0.\label{gtb4}
\ee
Eq. (\ref{gtb3}) can be integrated to obtain
\be
k^3=\frac{c_{\phi}}{R^2}.
\label{cphi}
\ee

The equations for the expansion $\theta$ and the shear $\sx$ of a beam
take the form
\bea
&&\frac{d\theta}{d\lx}=-\frac{1}{4M^2}\rho 
\left(k^0\right)^2-{\theta^2}-\sigma^2
\label{exx1}\\
&&\frac{d\sigma}{d\lx}+2\theta\sigma=\frac{\left(k^3\right)^2R^2}{4 M^2}
\left(\rho-\frac{3\calm(r)}{4\pi R^3}\right),
\label{exx2}
\eea
while the equation for the beam area can be written as
\be
\frac{1}{\sqrt{A}}\frac{d^2\sqrt{A}}{d\lx^2}=
-\frac{1}{4M^2}\rho \left( k^0\right)^2  -\sigma^2.
\label{exx3}
\ee
It is clear from eq. (\ref{exx2}) that the shear is generated by 
inhomogeneities, for which the local energy density is different from
the average one. 

\subsection{Friedmann-Robertson-Walker (FRW) background}

The FRW metric is a special case of the LTB metric with 
\begin{eqnarray}
&&R(t,r)=a(t)r
~~~~~~~~~~~~~~~
f(r)=cr^2,~~~c=0,\pm 1
\label{tbfrw1} \\
&&\rho=\frac{c_\rho}{a^{3}(t)}
~~~~~~~~~~~~~~~~~~~~~~~
\calm(r)=\frac{4\pi}{3} c_\rho r^3.
\label{tbfrw2} \end{eqnarray}
The geodesic equations have the solution
\begin{eqnarray}
k^0=\frac{c_t}{a(t)} \label{gsolfrw1} \\
k^1=\pm\frac{(1+cr^2)^{1/2}}{a(t)^2}\left[c_t^2
-\frac{c^2_\phi}{r^2} \right]^{1/2}
\label{gsolfrw2} \\
k^3=\frac{c_\phi}{a^2(t)r^2}.
\label{gsolfrw3} \end{eqnarray}
The shear can be consistently set to zero as the r.h.s. of eq. (\ref{exx2})
vanishes.

The equation for the beam area (\ref{exx3}) is most easily solved
if the center of the coordinate system is taken at the location of the
beam source. This implies that $c_\phi=0$ in eq. (\ref{gsolfrw3}).
It can be easily checked that the solution of eq. (\ref{exx3}) for an outgoing
beam is
\be
A(\lx) = r^2(\lx)\, a^2(t(\lx)) \, \Omega_s.
\label{asol} \ee
The constant $\Omega_s$ can be identified with the solid angle spanned by a 
certain beam when the light is emitted by a point-like isotropic source. 
We point out that, according to eq. (\ref{tha}), 
the beam expansion $\theta$ diverges at the
location of the point-like source ($r=0$).

We also obtain 
\be
\frac{d\sqrt{A}}{d\lx} =
k^1\,a\sqrt{\Omega_s}
+r\,\dot{a}k^0\sqrt{\Omega_s}.
\label{dsqa} \ee
At the location of the source the second term vanishes, so that 
\be
\left. \frac{d\sqrt{A}}{d\lx} \right|_{r=0}=
k^1(0)\,a(t_s)\sqrt{\Omega_s}=\frac{dr}{d\lx}(0)\,a(t_s)\sqrt{\Omega_s}
=\frac{dt}{d\lx}(0)\,\sqrt{\Omega_s}.
\label{dsqa0} \ee
If we normalize the scale factor so that $a(t_s)=1$ at the time
of the beam emission, we recover the standard expression 
$A=r^2 \Omega_s$ in flat space-time. 
In the following sections we study light propagation in more general
backgrounds. We assume that the light emission near the source is not
affected by the large scale geometry. By choosing an affine parameter
that is locally $\lx=t$ in the vicinity of the source, we can set
\be
\left. \frac{d\sqrt{A}}{d\lx} \right|_{\lx=0}=\sqrt{\Omega_s}.
\label{init1} \ee
This expression, along with
\be
\left. \sqrt{A} \right|_{\lx=0}=0,
\label{init2} \ee
provide the initial conditions for the solution of eq. (\ref{sqrta}). 
For isotropic sources, we also expect the beam shear to vanish at the
time of emission.

\subsection{Schwarzschild background}

The metric has the form
\begin{equation}
ds^{2}=-\left(1-\frac{r_s}{r} \right) dt^2
+\left(1-\frac{r_s}{r} \right)^{-1}dr^2+r^2 d\Omega^2,
\label{metricsch}
\end{equation}
where $d\Omega^2$ is the metric on a two-sphere and 
$r_s= \calm_0/(8\pi M^2)$ the Schwarzschild radius.

Without loss of generality we consider 
geodesic null curves on the plane with $\theta=\pi/2$. 
The geodesic equations are ($k^i={dx^i}/{d\lambda}$)
\bea
&&\frac{dk^0}{d\lx}+\frac{r_s}{r^2}
\left(1-\frac{r_s}{r} \right)^{-1}k^0 k^1
=0 \label{gsch1}\\
&&\frac{dk^1}{d\lx}+\left[
\frac{r_s}{2}-r\left(1-\frac{r_s}{r} \right)
\right]\left( k^3 \right)^2
=0\label{gsch2}\\
&&\frac{dk^3}{d\lx}+\frac{2}{r}k^1 k^3
=0\label{gsch3}\eea
and the null condition
\be
-\left(1-\frac{r_s}{r} \right) \left( k^0 \right)^2
+\left(1-\frac{r_s}{r} \right)^{-1}\left(k^1\right)^2+r^2 \left( k^3\right)^2
=0.\label{gsch4}
\ee
Their solution is
\begin{eqnarray}
&&k^0=c_t\left( 1-\frac{r_s}{r} \right)^{-1}
\label{gsolsch1} \\
&&k^1=\pm\left[c_t^2-\frac{c^2_\phi}{r^2} 
\left(1-\frac{r_s}{r} \right)\right]^{1/2}
\label{gsolsch2} \\
&&k^3=\frac{c_\phi}{r^2}.
\label{gsolsch3} \end{eqnarray}

The equations for the expansion $\theta$ and the shear $\sx$ of a beam
take the form
\bea
&&\frac{d\theta}{d\lx}=-{\theta^2}-\sigma^2
\label{exs1}\\
&&\frac{d\sigma}{d\lx}+2\theta\sigma=-\frac{3\left(k^3\right)^2}{2}
\frac{r_s}{r},
\label{exs2}
\eea
while the equation for the beam area can be written as
\be
\frac{1}{\sqrt{A}}\frac{d^2\sqrt{A}}{d\lx^2}=
-\sigma^2.
\label{exs3}
\ee

\subsection{Luminosity distance and redshift}

In order to define the luminosity distance, we consider photons
emitted within a solid angle $\Omega_s$
by an isotropic source with luminosity $L$.
These photons are detected 
by an observer for whom the light beam
has a cross-section $A_o$. 
The redshift factor is
\be
1+z=\frac{\omega_s}{\omega_o}=\frac{k^0_s}{k^0_o},
\label{redshiftt} \ee
because the frequencies measured at the source and at the observation point 
are proportional to the values of $k^0$ at these points.
The energy flux $f_o$ measured by the observer is
\be
f_o=  \frac{L}{4\pi D_L^2}= \frac{L}{4\pi}
\frac{\Omega_s}{(1+z)^2 A_o}.
\label{lumm} \ee
The above expression allows the determination of the luminosity distance
$D_L$ as 
a function of the redshift $z$. The beam area can be calculated by solving 
eq. (\ref{sqrta}), with initial conditions given by eqs. 
(\ref{init1}), (\ref{init2}), while the redshift is given by 
eq. (\ref{redshiftt}).

In the case of a FRW background, substitution of eq. (\ref{asol})
in eq. (\ref{lumm}) gives the standard expression
$D_L^2 = (1+z)^2 r^2_o a^2_o$, where $r_o$ corresponds to the 
comoving radial coordinate of the observer (for the source
$r_s=0$), and $a_o$ to the value of
the scale factor at the time of detection of the light 
signal.

\section{Spherical collapse in a LTB background}
\setcounter{equation}{0}

The inhomogeneous cosmology we consider is a variation of the 
Swiss-cheese model \cite{swiss}. Within a homogeneous
background, described by a FRW metric, 
we consider spherical inhomogeneous regions. The metric 
within each region has a spherical symmetry around its center.
In the traditional scenario, the total mass within every region is
assumed to be concentrated at the center, so that the relevant metric
is the Schwarzschild one. It is well known that the two metrics (FRW and
Schwarzschild) can me matched on their common boundary without
the appearance of a singular energy density there. The mass parameter of
the Schwarzschild metric must be equal to the total energy 
within a homogeneous
region of radius equal to that of the inhomogeneity.
The study of light propagation in such a background was pioneered by 
Kantowski \cite{kantowski} and underlies the broad subject
of gravitational lensing \cite{lensing}. 

We are interested in light propagation in a background that accounts for
the process of gravitational collapse. Our main aim is to examine how the
luminosity-redshift relation is modified when the light crosses regions
within which the deviation of the matter distribution 
from homogeneity is time-dependent.
Within a FRW background we consider spherical regions where the
space-time is described by the LTB metric (\ref{metrictb}). 
The assumption
of spherical symmetry makes the problem tractable. In a realistic scenario we
expect deviations from exact spherical symmetry, but the essense of the 
conclusions should remain unaffected.

The choice of the two
arbitrary functions $\calm(r)$ and $f(r)$ in eq. (\ref{metrictb})
can lead to different physical situations.  
The mass function $\calm(r)$ is related to the initial matter distribution. 
The function $f(r)$ defines an effective curvature term 
in eq. (\ref{tb1}). 
We can also interpret $f(r)$ as part of the initial radial velocity of the
fluid. 
We work in a gauge in which $R(0,r)=r$. We parametrize the initial 
energy density as $\rho_i(r)=\left( 1+\ex(r)\right)\rho_{0,i}$, with
$\rho_i(r)=\rho(0,r)$ and $|\ex(r)| < 1$.
The initial energy density of the homogeneous background is
$\rho_{0,i}=\rho_{0}(0)$. If the size of the inhomogeneity is 
$r_0$, a consistent solution requires 
$4\pi\int_0^{r_0} r^2 \ex(r) dr=0$, 
so that 
\be
\calm(r_0)=4\pi\int_0^{r_0} 
r^2\rho(r)\, dr=\frac{4\pi}{3} r^3_0 \rho_{0,i}.
\label{req} \ee

We assume that 
at the initial time $t_i=0$ the expansion rate $H_i=\dot{R}/R=\dot{R}'/R'$
is given for all $r$ by the standard
expression in homogeneous cosmology:
$H_i^2=\rho_{0,i}/(6M^2)$.
Then, eq. (\ref{tb1}) with $R(0,r)=r$ implies that 
\be
f(r)=\frac{\rho_{0,i}}{6M^2}r^2\left( 
1-\frac{3\calm(r)}{4\pi r^3 \rho_{0,i}}\right).
\label{fr} \ee
The spatial curvature of the LTB geometry is 
\be
^{(3)}R(r,t)=-2 \frac{(fR)'}{R^2 R'}.
\label{spcurv} \ee
For our choice of $f(r)$ we find that at the initial time 
\be
^{(3)}R(r,0)=-6 H^2_i
\left(1- \frac{\calm'}{4\pi  r^2 \rho_{0,i}} \right)
=-6H^2_i
\left( 1-\frac{\rho_i(r)}{\rho_{0,i}} \right).
\label{spcurvin} \ee
Overdense regions have positive spatial curvature, while underdense ones
negative curvature. 
This is very similar to the initial condition considered 
in the model of spherical collapse \cite{sphc1}.

When the inhomogeneity is denser near the center, we have
$f(r) < 0$ for $r < r_0$ and $f(r)=0$ for $r \geq r_0$.
It is then clear from eq. (\ref{tb1}) that, in an 
expanding Universe with 
increasing $R$, the central region will have $\dot{R}=0$ at some point in 
its evolution and will stop expanding. Subsequently, it will reverse its
motion and start collapsing. 

This point becomes clearer if we consider an example with 
$\ex(r)=\ex_1>0$ for $r<r_m$, and 
$\ex(r)=-\ex_1 r_m^3 /\left(r^3_0-r^3_m \right)$ for $r_m<r<r_0$.
In the region $r<r_m$ we have $R(t,r)=a(t)r$,
$f(r)=-\left(\ex_1\rho_{0,i}/6M^2 \right) r^2$ and 
\be
\left(\frac{\dot{a}}{a} \right)^2=\frac{1}{6M^2}
\frac{(1+\ex_1)\rho_{0,i}}{a^3}-\frac{\ex_1 \rho_{0,i}}{6M^2}\frac{1}{a^2}.
\label{stand} \ee
The evolution is typical of a closed homogeneous Universe with curvature
proportional to $\ex_1$.
For $r>r_m$, the effective curvature term
\be
f(r)=-\frac{\ex_1 \rho_{0,i}}{6M^2}\frac{r_m^3}{r}
\left( 
\frac{r^3_0-r^3}{r_0^3-r^3_m}
\right)
\label{curvf} \ee
remains negative, but goes to zero with $r\to r_0$. The shells with 
$r_m<r<r_0$ stop expanding and eventually collapse, but at progressively later
times. The shell with $r=r_0$ expands forever. 

We can obtain an analytical expression for the growth of the central
overdensity in this simple 
model. We define the quantity 
\be
\zeta(t)=\frac{R(t,r_0)/r_0}{R(t,r_m)/r_m}-1.
\label{zeta} \ee
For $0<\ex_1 \ll 1$, the ratio of the 
energy density within the perturbation to the energy density 
far from it is given by the factor 
$(1+\zeta)^3$.
For 
$\zeta \lta 1$ and $t\gta H^{-1}_i$ we find
\be
\zeta \simeq \frac{\ex_1}{5} 
\left(\frac{3}{2} H_i t\right)^{2/3}.
\label{zetat} \ee
At a time $t_2\simeq \ex_1^{-3/2}H^{-1}_i$ we have $(1+\zeta)^3\simeq 2$, which
means that 
the energy density within the perturbation is twice the
value in the homogeneous region. 
The phase of gravitational collapse starts at a time 
$t_c \simeq 1.5\, t_2$.
The growth of the perturbation $\sim t^{2/3}$ 
is in qualitative agreement with
the behaviour predicted by the Jeans analysis for adiabatic
subhorizon perturbations
in a matter dominated Universe.
Our model can be viewed as an exact solution of the Einstein
equations that is consistent with the behaviour expected from perturbation
theory in the region of its applicability.  
It also has the same qualitative features as the model of spherical collapse
\cite{sphc1}.

This simple model has a physical interpretation for $\ex_1 < 0$ as well.
In this case the central region is an underdensity, surrounded by a shell
with density larger than the average. 
The approximate analytical treatment remains the same as before. The effective
curvature term is always positive and goes to zero for $r\to r_0$.
The central region expands faster than the surrounding shell and its density
drops faster than the average. On the other hand, the relative shell size
shrinks and its energy density grows relative to the average density.
We shall study a model with this typical behaviour
in the following.

We emphasize that our choice of the functions $\calm(r)$ and $f(r)$ is 
not the only one possible. A general solution of the 
effective Friedmann equation (\ref{tb1})
also depends on an arbitrary 
function $t_B(r)$ that determines the local Big Bang time.
This results from the fact that eq. (\ref{tb1})
can be integrated with respect to $t$ for fixed $r$.
The solution
depends on the combination $t-t_B(R)$, with $t_B(r)$ arbitrary. 
In our model we have set $t_B(r)=0$. An example of a study of structure
formation, exploring the full freedom of the LTB metric, is given in ref.
\cite{hellaby}.

One final remark concerns the form of the functions 
$\rho_i(r)$ and $\ex(r)$. If they are discontinuous (as in our simple example)
the functions $\calm'(r)$, $f'(r)$ and $R'(t,r)$ become discontinuous as well. 
This implies that the metric (\ref{metrictb}), (\ref{brttb}) also becomes 
discontinuous. Of course the discontinuity of $\rho_i(r)$ is only an 
approximation. The realistic 
physical situation involves a fast variation of the
density. However, in order to avoid complications in the interpretation of
our results, we always choose continuous functions $\rho_i(r)$ and $\ex(r)$
in the following.

\section{The model}
\setcounter{equation}{0}

Our expressions simplify if we switch to dimensionless variables.
We define $\ttb=t H_i$, $\rb=r/r_0$, $\Rb=R/r_0$, where 
$H^2_i=\rho_{0,i}/(6M^2)$ is the initial homogeneous
expansion rate and $r_0$ gives the size of the inhomogeneity in
comoving coordinates.
The evolution equation becomes
\be
\frac{\dot{\Rb}^2}{\Rb^2}=\frac{3\calmb(\rb)}{4\pi\Rb^3}+
\frac{\fb(\rb)}{\Rb^2},
\label{eind} \ee 
with $\bar{\calm}=\calm/(\rho_{0,i}r^3_0)$
and $\fb=6M^2f/(\rho_{0,i}r_0^2)=f/\Hb^2_i$, $\Hb_i=H_ir_0$.
The dot now denotes a derivative with respect to $\ttb$.

\begin{figure}[t]
\includegraphics[width=11cm, angle=0]{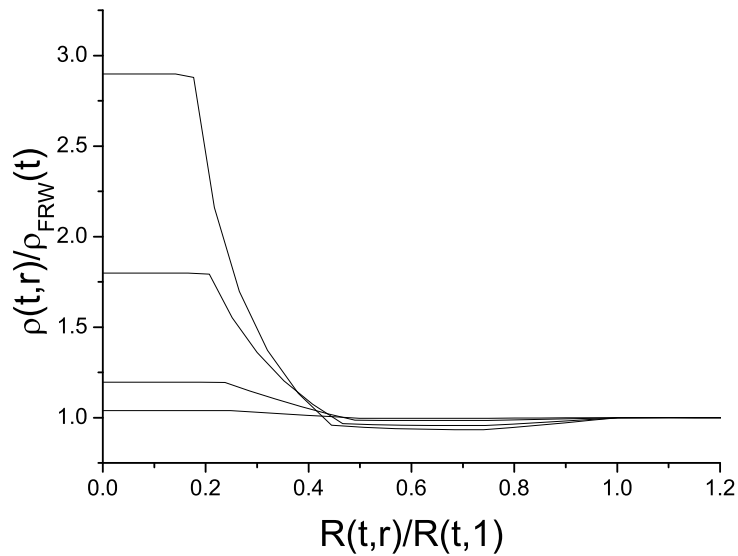}
 \caption{\it
The evolution of the density profile for a central overdensity surrounded by
an underdensity.
}
 \label{fig1}
 \end{figure}

\begin{figure}[t]
\includegraphics[width=11cm, angle=0]{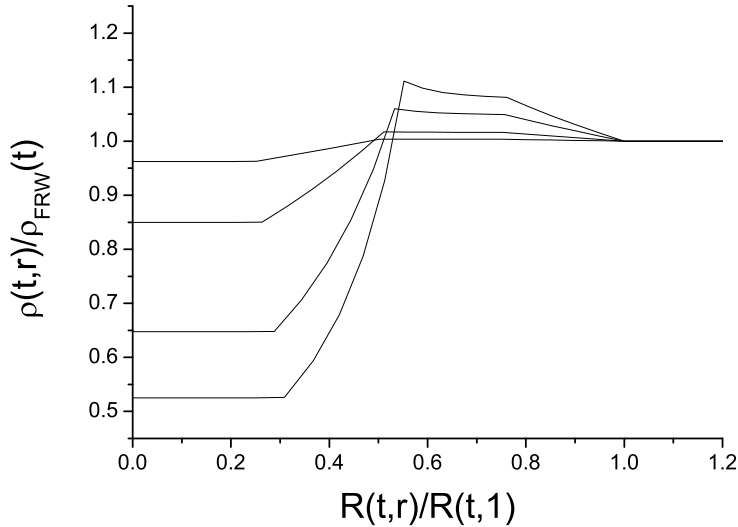}
 \caption{\it
The evolution of the density profile for a central underdensity surrounded by
an overdensity.
}
 \label{fig2}
 \end{figure}

The typical cosmological evolution in our model is displayed in figs. 
\ref{fig1} and \ref{fig2}. The initial density
$\rhb_i(r)=\rho_i(r)/\rho_0=1+\ex(r)$
is constant $\rhb_i=1+\ex_1$ in the region $\rb\leq 0.25$, 
constant $\rhb_i=1+\ex_2$ in the region $0.5\leq \rb \leq 0.75$, and
$\rhb_i=1$ for $\rb \geq 1$. In the intervals $0.25 \leq \rb\leq 0.5$
and $0.75 \leq \rb \leq 1$ it interpolates linearly between the values
at the boundaries. For fig. \ref{fig1} we take $\ex_1=0.01$, while
for fig. \ref{fig2} we use $\ex_1=-0.01$. The values of $\ex_2$ are 
fixed by the requirement that $\int_0^1\ex(r)r^2 dr=0$.

In fig. \ref{fig1} we show the typical behaviour in the case of 
a central overdensity that is surrounded by an underdense region.
We display the density profile at times $\ttb=10,100,500,1000$.
We normalize the energy density to that of a homogenous FRW background
(given by $\rhb_{FRW}(t)=\rhb(t,1)$). The initial time corresponds to the
curve with the smallest deviation from 1, while the final to the curve
with the largest deviation.
We observe that 
the density contrast grows and eventually becomes ${\cal O}(1)$. 
The central region becomes denser with time, and the underdense region
emptier. The central region will stop expanding at a time 
$\ttb \sim 1500$, reverse its motion and collapse towards the center.
The same will happen to all the outer shells, but at progressively later
times. The shell with $\rb=1$ expands forever. One important feature is
that the relative radius of the central region becomes smaller with time,
as the expansion is slower than the average within this region.

In fig. \ref{fig2} we display the typical behaviour in the case of 
an underdensity surrounded by an overdense region. 
We display the density profile at times $\ttb=10,100,500,1000$. 
As before, the deviation from 1 is smallest at the initial time and largest
at the final. The 
central energy density drops relative to the homogeneous background, while
the surrounding region becomes denser.
The radius of the central underdensity grows relative to the total size 
of the inhomogeneity, as this region expands faster than the average.

\section{Light propagation and luminosity distance}

We are interested in light propagation in the gravitational background
of the configuration described in the previous section. 
We take the affine parameter $\lx$ to have the dimension of time and
we define the dimensionless variables 
$\lb=H_i\lx $,
$\kb^0=k^0$, $\kb^1=k^1/\Hb_i$, $\kb^3=r_0 k^3$.
The geodesic equations (\ref{gtb1})--(\ref{gtb4}) maintain their form, with
the various quantities replaced by the barred ones, and the combination
$1+f$ replaced by $\Hb_i^{-2}+\fb$. For geodesics going through
subhorizon perturbations with 
$\Hb_i\ll 1$ the effective curvature term $\fb$ plays a minor roll. However,
this term is always
important for the evolution of the perturbations, as can be seen from
eq. (\ref{eind}).

We also define the dimensionless expansion $\thb=\theta/H_i$ and
shear $\sxb=\sx/H_i$. The optical equations (\ref{exx1})--(\ref{exx3})
take the form
\bea
&&\frac{d\thb}{d\lb}=-\frac{3}{2}\rhb 
\left(\kb^0\right)^2-{\thb^2}-\sxb^2
\label{exx1r}\\
&&\frac{d\sxb}{d\lb}+2\thb\sxb=\frac{3}{2}
\left(\kb^3\right)^2\Rb^2
\left(\rhb-\frac{3\calmb(\rb)}{4\pi \Rb^3}\right),
\label{exx2r}\\
&&\frac{1}{\sqrt{\Ab}}\frac{d^2\sqrt{\Ab}}{d\lb^2}=
-\frac{3}{2}\rhb \left( \kb^0\right)^2  -\sxb^2,
\label{exx3r}
\eea
with $\rhb=\rho/\rho_{0,i}$.
The initial conditions (\ref{init1}), (\ref{init2}) become
\bea
&&\left. \frac{d\sqrt{\Ab}}{d\lb} \right|_{\lb=0}=\frac{1}{\Hb_i}\sqrt{\Omb_s}
=\sqrt{\Omega_s}.
\label{init1r} \\
&&\left. \sqrt{\Ab} \right|_{\lx=0}=0,
\label{init2r} \eea
with $\Ab=H_i^2A$ and $\Omb=\Hb_i^2 \Omega$.

\begin{figure}[t]
\includegraphics[width=11cm, angle=0]{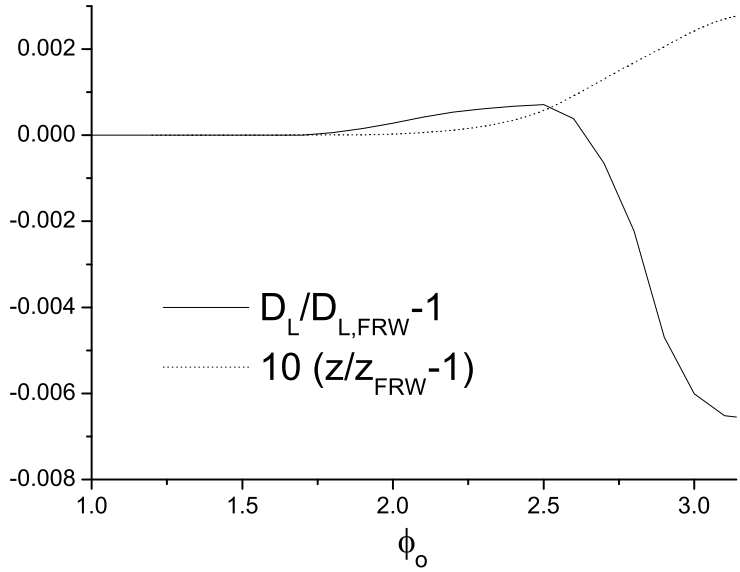}
 \caption{\it
The luminosity distance $D_L$, relative to the case of homogeneous 
cosmology, for light beams crossing a configuration with a central
overdensity and exiting at various angles $\phi_o$.}
 \label{fig3}
 \end{figure}

For our study we integrate the geodesic and optical equations numerically in 
the backgrounds described in the previous section.
In order to get a feeling of the qualitative effect on the luminosity 
distance, we consider first a light beam that crosses one spherical
inhomogeneous region. We assume that the light is emitted at a radius
$\rb=1.5$. Its trajectory is obtained by integrating eqs. 
(\ref{gtb1}), (\ref{gtb3}), (\ref{gtb4}) (in their rescaled form). 
Eq. \ref{gtb2}) is then
automatically satisfied. 
The initial value of $\bar{k}_0$ is arbitrary, as it corresponds to the
initial frequency of the emitted light.
The integration of eq. (\ref{gtb3}) gives
eq. (\ref{cphi}). The angle of entry of the beam into the inhomogeneous
region can be varied through the choice of $c_\phi$, or its
rescaled version $\bar{c}_\phi=c_\phi/r_0$. For $c_\phi=0$ the trajectory
goes through the center of the spherical configuration.

The form of the various light trajectories is not particularly 
illuminating. The important effect is the difference in light propagation
within an inhomogeneous background and a homogeneous one. In order to make
this comparison we consider two light trajectories (one in each background)
that start at a common point with $\rb=1.5$ and eventually approach another
common point with $\rb=1.5$. The value of $\bar{c}_\phi$ and the length of 
trajectories in terms of the affine parameter or the coordinate time 
are different. However, both light beams are emitted 
and received by comoving observers in the homogeneous region, with
the same spatial coordinates. The difference
in beam area and redshift is generated by the background that the 
beam crosses. Alternatively, one could choose to compare beams emitted
at the same point within the homogeneous region, with the same 
$c_\phi$. These would approach $\rb=1.5$ at different values of the
angular coordinate $\phi$. This choice gives very similar results to the
ones we describe.

\begin{figure}[t]
\includegraphics[width=11cm, angle=0]{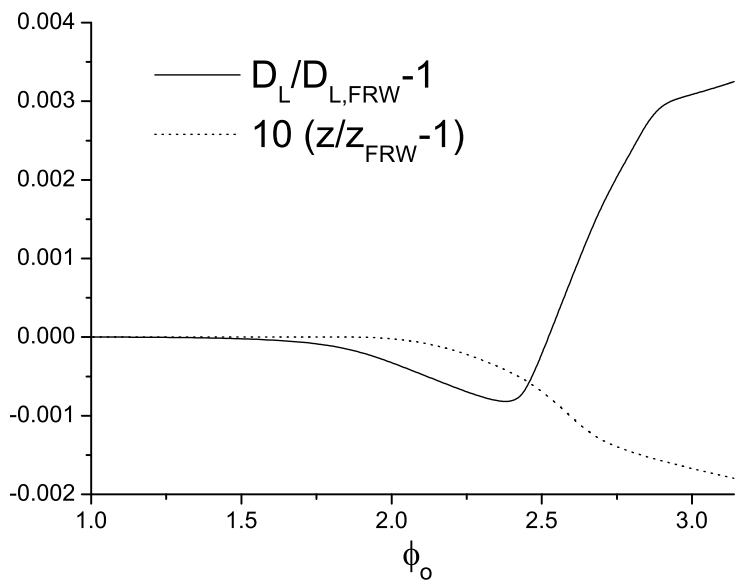}
 \caption{\it
The luminosity distance $D_L$, relative to the case of homogeneous 
cosmology, for light beams crossing a configuration with a central
underdensity and exiting at various angles $\phi_o$.}
 \label{fig4}
 \end{figure}

In figs. \ref{fig3}, \ref{fig4} 
we display the difference in luminosity distance and
redshift between inhomogeneous and homogeneous backgrounds, as 
a function of the comoving angle $\phi_o$ of the observer. 
The light is emitted from the point $\rb_s=1.5$, $\phi_s=0$, 
and observed at the point $\rb_o=1.5$, $\phi_o$. We have used
$\Hb_i=1$ and taken the emission time of the beam to be
$\ttb_s=550$. The background within which the light propagates is
not consistent with the theory of structure formation, 
as it includes a perturbation of horizon size 
with $|\ex_1|=0.01$, instead of the standard
$|\ex_1|=10^{-5}$. We choose such a background in order to 
make its qualitative effect on the beam properties more visible.
In the following section, in which we focus on a quantitative estimate
of the effect, we consider realistic backgrounds.

For the results displayed in fig. \ref{fig3} the 
inhomogeneity consists of a central overdensity surrounded by an
underdensity. For small values of $\phi_o$ the light trajectory stays within
the homogeneous region $1\leq \rb \leq 1.5$. As a result no difference
is observed. For larger angles the trajectory crosses only the underdense
part of the inhomogeneity. The resulting luminosity distance is larger 
than the one for a homogeneous background. At even larger angles the beam
enters the central overdense region. The luminosity distance becomes
smaller than the one in a homogeneous background. This behaviour is a
consequence of the term $\sim \rho (k^0)^2$ in the r.h.s. of 
eq. (\ref{exx3}).
If the effect is averaged over angles, a decrease of
the luminosity distance is expected.

The shear $\sx$ has only a minor effect in the evolution of $\sqrt{A}$
and the determination of the luminosity distance. The reason is that the
shear is generated by the difference of the local energy density $\rho$ and
an average density $\rho_{av}=\calm/(4\pi R^3/3)$. These coincide within
the central homogeneous region.
They are different in the surrounding underdense region. However, the later
gives an effect only for beams that stay within the overdensity a
short time, without significant 
modification of the respective luminosity distance. A significant effect can
be generated if the mass of the overdensity is concentrated near the 
center. In this case, eqs. (\ref{exx1})-(\ref{exx3}) give an effect 
similar to that of eqs. (\ref{exs1})-(\ref{exs3}) for a Schwarzschild 
geometry, with the additional influence of the expansion. It is known
that for a Schwarzschild geometry the shear plays a minor role, unless the
beam approaches the center at distances comparable to the Schwarzschild radius
\cite{kantowski}.

The relative difference in redshift is more than an order of magnitude
smaller than the respective difference in luminosity distance.
The reason is that the redshift, as opposed to the beam area, 
receives compensating contributions 
when entering and exiting the inhomogeneity. The net effect depends only
on the evolution with time of the inhomogeneity (the reflection of
the Rees-Sciama effect \cite{rees} 
in our formalism), and would be zero for a static one. 
We conclude that it is a good approximation to assume that the redshift 
remains unaffected by the presence of the inhomogeneity. This conclusion
is consistent with the results of the study based on the standard Swiss-cheese
model, in which the inhomogeneous region is modelled through the
Schwarzschild geometry \cite{kantowski}. 

In fig. \ref{fig4} we display the modification of the luminosity distance
and redshift in the case of central underdensity surrounded by an
overdensity. Again, we have used
$\Hb_i=1$ and taken the emission time of the beam to be
$\ttb_s=550$. We observe an effect on the luminosity distance that
is opposite to that in the previous case of a central overdensity.
Again, the modification of the redshift can be neglected in a good
approximation.
 
The comparison of figs. \ref{fig3} and \ref{fig4} demonstrates that 
the average modification of the luminosity distance, for a large
statistical sample of light beams crossing the
inhomogeneity at various angles, depends on the form of the 
configuration. 
The presence of underdense central regions, consistent with the 
appearance of large, approximately spherical voids in the matter distribution 
of the Universe, 
is expected to lead to an increase of the average luminosity distance 
relative to the homogeneous case. The opposite is expected if
the central regions are overdense. It must be emphasized that the average 
density is equal to that of the homogeneous case for both configurations
displayed in figs. \ref{fig1} and \ref{fig2}. In this sense, these
configurations do not
have to appear in equal numbers in a realistic model of the Universe. It
seems reasonable to suggest that the
dominance of voids, as deduced from observations \cite{peebles}, can be
linked with increased luminosity distances.

\section{The magnitude of the effect}

\begin{figure}[t]
\includegraphics[width=11cm, angle=0]{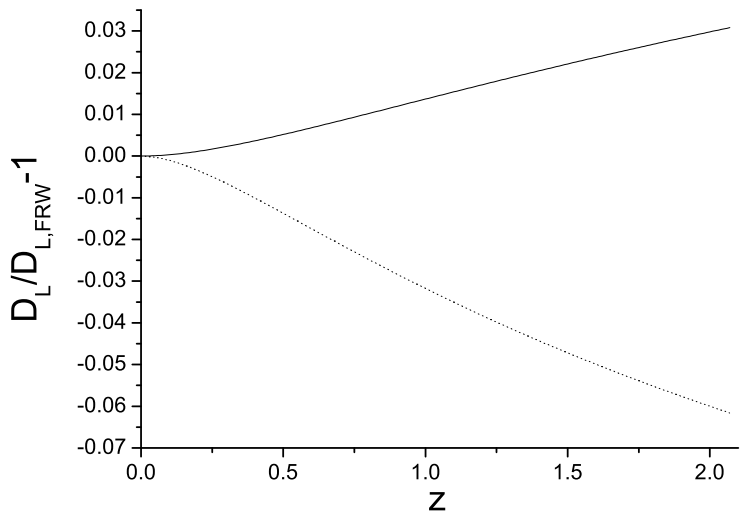}
 \caption{\it
The luminosity distance $D_L$ relative to the case of homogeneous 
cosmology, as a function of the redshift $z$, 
for light beams crossing diametrically several configurations with 
a central
underdensity (upper curve) or overdensity (lower curve).
}
 \label{fig5}
 \end{figure}

The important question is whether the dominance of configurations with 
central underdensities can induce an effect that would explain the
observed luminosity curves for distant supernovae. Our model is too
basic to address this question in full detail. However, we can use it
in order to estimate the magnitude of the expected effect in the real
Universe. 

The biggest increase in the luminosity distance relative to a homogeneous
cosmology is obtained if the Universe includes only void-like 
inhomogeneities with central
underdensities. The effect is maximized if the light is assumed to
cross these configurations passing through their center. The effect of
the evolution of the inhomogeneities must be taken into account, with
initial conditions consistent with the CMB spectrum (amplitude 
${\cal O}(10^{-5})$ at horizon crossing) and the observed matter
distribution today. Within our model, this can be achieved by 
using as initial condition a configuration with $\ex_1=-0.01$ and 
size relative to the horizon
$r_0H_i=\Hb_i=1/10$ at some initial time $\ttb_i=0$. This 
evolves similarly to fig. \ref{fig2} into a configuration with a
density contrast ${\cal O}(1)$ at a time $\ttb_f\simeq 1436$, which
we identify with the present. The size of the inhomogeneity relative to the
horizon today is
$R(t_f,r_0)H_f=\dot{R}(t_f,r_0)=\dot{\Rb}(\ttb_f,1) \Hb_i$. 
Our solution gives 
$\dot{\Rb}(\ttb_f,1)\simeq 0.076$, so that the size of the inhomogeneity 
becomes $R(t_f,r_0)\simeq 23 \,h^{-1}$ Mpc. This is of the order of the 
typical size of voids today. 
At times before the initial moment $\ttb_i$
from which we follow the evolution,
the inhomogeneity is only a small perturbation, 
with amplitude ${\cal O}(10^{-5})$ at horizon crossing.

We consider light beams that pass repeatedly through the inhomogeneities
we described above. The light is emitted at some time $\tb_s$ from a point
with $\rb=1.5$ within the homogeneous region. The initial
conditions for the beam area are given by eqs. (\ref{init1}), (\ref{init2}). 
We assume that the light moves radially
towards the center of the inhomogeneity (that has a fixed radius $\rb_0=1$ in
comoving coordinates), exits from the opposite
side and finally arrives at a point with $\rb=1.5$. 
Subsequently, the beam crosses the following inhomogeneity in a similar
fashion. The initial conditions are now set by the values of 
$\sqrt{\Ab}$ and $d\sqrt{\Ab}/{d\lb}$ at the end of the first crossing. 
Our assumption that the motion is radial produces the maximum
effect. In general the crossing should take place at a random angle, with a 
smaller total increase of the luminosity distance.
Of course, as time passes the profile of the inhomogeneities changes, as
depicted in fig. \ref{fig2}. 

The total number of crossings determines the total redshift of the beam
and the final beam area, related to the luminosity distance. We repeat
the calculation for several starting times that result in 
a variation of the total number of crossings. In fig. \ref{fig5}
(upper curve) we depict the resulting increase in the luminosity distance
relative to the homogeneous case, as a function of the redshift.
As we discussed in the previous section, the value of the redshift is
essentially unaffected by the presence of the inhomogeneities. (The effect
is smaller by more than an order of magnitude compared to the modification
of the luminosity distance.) In fig. \ref{fig5} we also depict the form
of the luminosity curve if the inhomogeneity has a central overdensity 
(lower curve). In this case the luminosity distance is decreased relative
to the homogeneous case.

It is clear from fig. \ref{fig5} that the influence of inhomogeneities on
the luminosity distance is a small effect, at most of the order of a few \%.
This must be compared to the required increase in the luminosity distance
in order to explain the supernova data, which is of the order of 30\% at
redshifts around 1.
This conclusion remains valid for other profiles of the inhomogeneities, as
long as the essential phenomenological requirements are satisfied 
(consistency with the observed large scale structure today, amplitude
${\cal O}(10^{-5})$ at horizon crossing).

It must be pointed out that the region of validity of our model 
does not extend to very high redshifts. Despite the fact that the 
amplitude of the perturbation goes to zero at early times in our model,
a significant increase of the luminosity distance is predicted even
for large $z$. We do not expect such a phenomenon to prevail in a model
that preserves the full spectrum of perturbations, instead of keeping only
one with a characteristic scale, as we did. 

Obtaining an analytical estimate of the modification of
the luminosity distance is not easy. However, an 
absolute upper bound can be derived from eq. (\ref{exx3}).
The ``focusing'' of a beam is minimized if the shear is negligible and
the energy density of the cosmological medium is set to zero in this 
equation. In our model this idealized situation
can be achieved if the central underdense regions of the inhomogeneities
become totally empty after a long evolution, 
while overdense spherical thin shells develop
around them. The effect of a shell, no matter how thin, on the beam area 
is not necessarily negligible. However, if we want
to derive only an upper bound on the increase of the luminosity distance, we
can set the energy density arbitrarily to zero in the optical equation only. 
Of course, the energy density still drives the cosmological
expansion through eq. (\ref{tb1}). 
The scale of the overall expansion is 
determined by the average energy density. In our model this is obvious
from the expansion in the homogeneous regions outside the inhomogeneities.

The equation that we derive in this way corresponds to the 
partially filled beam equation in FRW cosmology \cite{dyer,kantowski2},
if all the matter is assumed to be concentrated in 
dense objects that are not crossed by the light trajectories.
The same equation results from the 
optical equation (\ref{exs3}) within the holes of the standard Swiss-cheese
model if the shear is neglected. 
The luminosity distance as a function of the redshift can be derived
analytically in this case, assuming that the background expansion is
given by the standard Friedmann equation involving the average density
\cite{bound}. When comparing it to the luminosity distance in a homogenous
matter-dominated FRW cosmology, we find
\be
\frac{D_L}{D_{L,FRW}}-1=\frac{1}{5}
\frac{(1+z)^2-(1+z)^{-3/2}}{1+z-(1+z)^{1/2}}-1.
\label{analyt} \ee
For redshifts near 1, this expression gives an increase in the
luminosity distance by around 10\%. This is the maximum effect we could
expect, if the expansion rate is governed by the average density
through the standard Friedmann equation.

\section{Summary and conclusions}

The purpose of this study has been to determine the modification of
the luminosity distance of astophysical objects, such as supernovae,
which is generated by the appearance of large scale inhomogeneities. 
We constructed a picture different from that of the standard Swiss-cheese
model. The inhomogeneities are still modelled as spherical regions within
a homogeneous background, but the matter inside them is not concentrated
in very dense objects at their center. Instead, it is continuously 
distributed in regions of density below or above the average one. 
We assumed spherical symmetry for each one of the inhomogeneous
configurations. The LTB metric gives the most general description within
each spherical region 
if the cosmological fluid is assumed to be pressureless.
Moreover, it provides an automatic matching with the intermediate
homogeneous regions, as the FRW metric is a special case of the LTB one. 
As a result, our model automatically becomes 
an exact solution of the Einstein equations.

One fundamental assumption in our study was that neither 
the light source nor the observer occupy a preferred position in the
Universe. For this reason we assumed that the light is emitted and received
within homogeneous regions, while it crosses one or many 
inhomogeneous ones along its path. We derived the necessary 
optical equations that
describe the propagation of a light beam within a LTB background. 

The cosmological evolution within each spherical region has many realistic
elements when compared with the Universe at scales around and above 
$10\,h^{-1}$ Mpc. 
Central overdense regions become denser with time, with
underdense spherical shells surrounding them. Central underdense
regions turn into voids, surrounded by massive shells.
The time evolution of these configurations is in qualitative agreement with
perturbation theory and the spherical collapse model. 

The question we addressed is whether the luminosity distance of 
a light source may increase in such a background. We found that this
is indeed possible. If the inhomogeneities involve a central underdensity
and an overdense outer shell, the luminosity distance is larger on the
average than if the matter was distributed homogeneously. 
The opposite happens if the inhomogeneity is denser in the 
middle, surrounded by an underdense shell. In this case
the luminosity distance is
smaller than in homogeneous cosmology. 
These results imply that 
a description of the Universe as being composed of large voids,
with the matter being concentrated in the intermediate regions,
could provide a basis for the explanation of the
supernova data without dark energy. 

The problem with the picture emerging from our model is quantitative:
The increase in the luminosity distance arising from void domination is too
small to account for the supernova data. The relative increase 
at a redshift around 1 is of the order of a few \%. Moroever,
an absolute upper bound of around 10\% can be derived. This is significantly
smaller than the required increase, which is around 30\%.

We mention at
this point that in our model the change in redshift because of the presence of 
inhomogeneities is smaller than the change in luminosity distance 
by more than an order of magnitude. As a result, we do not expect a large
Rees-Sciama effect \cite{rees} that would be in conflict with the CMB data.
It is also possible that the presence of voids could explain
the observed large angle CMB anomalies \cite{silk}.

The open question is whether an alternative model of
large scale structure could result in a larger effect. There is a variety
of metrics describing inhomogeneous cosmologies, so this 
cannot be excluded. 
In our modelling we have made some fundamental assumptions 
that limit the possibilities. These are: \\
a) The overall scale factor 
is determined by the average density through the standard Friedmann equation.\\
b) The observer does not occupy a preferred position in
the Universe, such as the center of a significant underdensity. In practice,
we studied light signals that originate in a homogeneous region, cross
a large number of inhomogeneities, and are detected by an observer within
a homogeneous region. \\
c) The inhomogeneities have a characteristic scale 
${\cal O} (10)\, h^{-1}$ Mpc today. Their evolution is consistent with
the standard theory of structure formation. In particular,
the density perturbations are
${\cal O}(10^{-5})$ at horizon crossing. \\
The first two assumptions lead to the LTB Swiss-cheese model that we studied.
The third one is consistent with the observed size of voids in the matter
distribution of the Universe.

If the first assumption is maintained we do not expect a significant 
increase of the luminosity distance, independently of the modelling of the
inhomogeneous regions.  
The analytical 
upper bound of approximately 10\% at redshifts around 1 would remain
valid in all such cosmologies, as long as the overall scale factor 
is determined by the average density through the standard Friedmann equation.

Of course, one could try to build a cosmology that deviates significantly from
the FRW one, so that the effective Friedmann equation receives large
corrections. An important limitation is that 
acceleration of the local volume expansion in the absence of vorticity
requires $\rho +3p < 0$, with 
$\rho$ the local energy density and $p$ the pressure \cite{seljak}. 

It has been suggested that the average expansion rate in a given background
may deviate from the local one, with the corrections accounting for the
observed acceleration \cite{rasanen1,rasanen2,buchert,romano}. 
However, how the averaging 
is reflected in the features of a light beam transmitted in the 
particular background is not clear yet.  
Obviously the light propagates within the exact local metric, for which
the optical equations lead to an unambiguous determination of the luminosity
distance. On the other hand, 
the notion of averaging becomes crucial for a complicated mass distribution,
for which the exact local metric cannot be determined.
In this case
the averaging should be a method of determining the
gross features of light propagation without considering all the
details of the exact metric. 
More work is needed in order to have a clear understanding of the 
connection between averaging and beam features.
For example, within our model the ``backreaction'' term
appearing in the averaged Raychaudhuri equation \cite{buchert} is zero. 
Despite that, we observe a modification of the luminosity distance.

The basic features required for the averaging to induce a significant 
deviation from the standard Friedmann expansion were determined
recently \cite{buchert2}. The average spatial curvature plays an
important role in this respect, as it is coupled to large spatial variances of
the local expansion rate and shear. Within our model the average 
spatial curvature is zero, so that this mechanism of ``backreaction'' through
averaging is absent.
This is a consequence of the initial conditions 
we chose for the LTB metric. As we discussed in section 3, our 
initial conditions guarantee that the evolution of the
inhomogeneities is consistent with the standard scenario of structure
formation (our third assumption). Maintaining this consistency 
in backgrounds with large average spatial curvature (similar to those
proposed in refs. \cite{enqvist,chuang})
is a challenging 
problem that merits further study in the future. An exact background
that preserves this consistency could help test the proposal
of ``backreaction'' through averaging, by studying the exact
light propagation in it.

Another possibility in order to explain the supernova data 
is to violate our second assumption and
place the observer in a special position in
the Universe. It has been observed that 
any form of the luminosity distance as a function of redshift can be 
reproduced for an observer at the center of 
the LTB metric \cite{mustapha}. The reason is that, 
within an anisotropic geometry, the volume expansion rate results from the  
averaging of unequal expansion rates along various directions. In particular,
for the LTB metric the radial expansion can be accelerating, while the
tangential one decelerating \cite{apo}. Of course, the local volume
expansion is always decelerating for $\rho+3p>0$. An observer located
at the center of the LTB metric receives signals only along the
radial direction, from which he may infer an accelerating expansion.

Reproducing the 
supernova data in LTB models requires a variation of the 
density or the expansion rate over distances 
${\cal O}(10^3)\,h^{-1}$ Mpc \cite{alnes}--\cite{biswas}
(also see \cite{tomita}).
This implies inhomogeneities at scales much larger than the ones
allowed by our third assumption. 
Moreover, in order to avoid a conflict with the isotropy of the 
CMB, the location
of the observer must be very close to 
the center of the spherical configuration 
described by the LTB metric.
Typically the required density contrast is of order ${\cal O}(1)$.
A notable exception is discussed in ref. \cite{enqvist}. In the 
model presented there, the matter 
distribution is homogeneous, but the expansion rate varies by roughly 
15\% over distances ${\cal O}(10^3)\,h^{-1}$ Mpc. This results from a 
variation of the spatial curvature. Even though the conflict with
the observed matter distribution is avoided in this way, 
the lack of consistency with the theory of structure formation and
the requirement of a preferred location of
the observer are the weak points of this approach. 

In all the models that reproduce the supernova data by placing the
observer at the center of the LTB geometry 
the Universe is very inhomogeneous at earlier times. The problem of 
initial conditions remains open, while the consistency with the amplitude of
the temperature fluctuations of the CMB is not obvious. In our approach we
constrained our modelling of the inhomogeneities so that it
leads to consistency with the theory of structure formation. Our findings
indicate that 
the constraints do not permit the reproduction of the supernova data 
without dark energy.

It is worth pointing out that the model we considered has zero vorticity. 
The presence of vorticity in the background geometry
gives a positive contribution to the expansion of the beam area
\cite{peebles}. Such models are difficult to study, but provide another
unexplored possibility for the explanation of the supernova
luminosity distance without dark energy.

As a final remark we mention that the effect we studied in this work
is important for the correct determination of the energy content of
the Universe. Even in the presence of dark energy, the corrections to
the luminosity distance arising from inhomogeneities must be taken into
account, as they can modify significantly the deduced contributions from
dark matter and dark energy to the total energy density \cite{kantowski2}.

\vspace {0.5cm}
\noindent{\bf Acknowledgments}\\
\noindent 
We would like to thank S. Rasanen for useful discussions and comments.
This work was supported by the research program 
``Pythagoras II'' (grant 70-03-7992) 
of the Greek Ministry of National Education, partially funded by the
European Union.

\vskip 1.5cm

\newpage

\section{Appendix A}
\setcounter{equation}{0}

For completeness, we derive in this appendix
the optical equations in a form that is 
convenient for our study. We follow closely the presentation in ref.
\cite{peebles}. 

The equation for the deviation between two neighboring null
geodesics $x^i(\lambda)$ and $x^i(\lambda)+\xi^i(\lambda)$ is
\be
\frac{D^2\xi^i}{d\lambda^2}=R^i_{~jkl}k^jk^k\xi^l.
\label{dev}
\ee
The symbol $D$ denotes a covariant derivative, while $\lambda$ is an 
affine parameter.
A null geodesic is defined through the relations
\be
k^i=\frac{dx^i}{d\lambda},~~~~~~~~~~~~~~~k^i_{~;\,j}k^j=0,~~~~~~~~~~~~~~~k^ik_i=0.
\label{null}
\ee

We are interested in finding a pseudo-orthonormal basis of vectors 
along the path of light, 
in which we can describe the 
evolution of the cross section of a beam of light. As such we use $k^i$, 
a null vector $w^i$ satisfying $w^i k_i=-1$, 
and $L^i_1, L^i_2$, which are space-like unit 
vectors, orthogonal to the light ray and to each other.
We require that all these be parallely propagated along the path, so as to
keep them orthogonal and normalized. In summary, we have along the path
\begin{eqnarray}
&&kk=ww=0,~~~~~~~~L_1L_1=L_2L_2=1,~~~~~~~~kw=-1,
\label{ortho1} \\
&&kL_1=kL_2=wL_1=wL_2=L_1L_2=0.
\label{ortho2} \end{eqnarray}

We can choose the initial deviation $\xi^i$ such that a freely moving 
observer sees it perpendicular to the light ray: $k_i\xi^i=0$, $w_i\xi^i=0$. 
It is easy then to check that the deviation remains orthogonal to the rest
of the path. As a result, 
$\xi^i$ can be expressed as a linear combination of the vectors $L^i_a$
\be
\xi^i(\lambda)=\sum_{a=1,2}\dx_a(\lambda)L^i_a,
\ee
where the scalars $\dx_a$ are the proper orthogonal components of 
the separation of the two light rays. 
Substituting this expansion into eq. (\ref{dev}) 
and using the orthogonality of $L^i_a$, we get
\be
\frac{d^2\dx_a}{d\lambda^2}=
\sum_bA_{ab}\dx_b,~~~~~~~~~~~A_{ab}=R_{ijkl}L^i_ak^jk^kL^l_b.
\ee
The decomposition of the curvature tensor into the Ricci and Weyl tensors
\bdm
R_{ijkl}=\frac{1}{2}\left(g_{ik}R_{jl}-g_{il}R_{jk}-g_{jk}R_{il}+g_{jl}R_{ik}\right)-\frac{1}{6}\left(g_{ik}g_{jl}-g_{il}g_{jk}\right)R+C_{ijkl}
\edm
gives
\be
A_{ab}=-\frac{1}{2}R_{ij}k^ik^j\delta_{ab}+C_{ijkl}L^i_ak^jk^kL^l_b.
\label{aab1} \ee

The components $\dx_a$ obey the equation
\be
\frac{d\dx_a}{d\lambda}=\sum_b\left(\theta\delta_{ab}+\sx_{ab}\right)\dx_b,
\label{gen} \ee
where $\theta$ is the expansion of the beam and $\sigma_{ab}$ 
the symmetric and traceless shear tensor
\be
\sigma_{ab}=\left( \begin{array}{cc}
\sx_1 & \sigma_2 \\
\sigma_2 & -\sx_1 \\
\end{array} \right).
\label{sxaba}
\ee
The later satisfies
\be
\sum_c \sx_{ac}\sx_{cb}=
\left(\sx_1^2+\sx_2^2\right) \delta_{ab}=\sx^2\delta_{ab}.
\ee
The second derivative of eq. (\ref{gen}) implies that
\be
A_{ab}=\left(\frac{d\theta}{d\lambda}+\theta^2+\sx^2\right)\delta_{ab}
+\left(\frac{d\sx_{ab}}{d\lambda}+2\theta \sx_{ab}\right).
\label{aab2} \ee
Using the identity 
$
\sum_{a}C_{ijkl}L^i_ak^jk^kL^l_a=0,
$
we can see that the second term of (\ref{aab1}) is traceless, 
as is the second term of (\ref{aab2}). We have then
\bea
\frac{d\theta}{d\lambda}+\theta^2+\sx^2=-\frac{1}{2}R_{ij}k^ik^j
\label{expr}\\
\frac{d\sx_{ab}}{d\lambda}+2\theta \sx_{ab}=C_{ijkl}L^i_ak^jk^kL^l_b.
\label{shr} \eea

The components $\dx_a$ at neighboring positions along the path are
\be
\bar{\dx}_a\equiv \dx_a(\lambda+\delta\lambda)=\dx_a(\lambda)+\left(\theta \dx_a
+\sum_b \sx_{ab}\dx_b\right)\delta\lambda.
\ee 
The proper cross-section area A of the beam at neighboring positions 
along the path is 
\be
A(\lambda+\delta\lambda)=\int d\bar{\dx}_1d\bar{\dx}_2=\int d\dx_1d\dx_2
\frac{\partial \bar{\dx}}{\partial \dx}
=\frac{\partial \bar{\dx}}{\partial \dx}A(\lambda).
\ee
Evaluating the Jacobian $\partial \bar{\dx}/\partial \dx$
results in 
\be
\frac{dA}{d\lambda}=2\theta A.
\label{thaa} \ee
Eqs. (\ref{expr}) and (\ref{shr}) can now be written as
\bea
&&\frac{1}{\sqrt{A}}\frac{d^2\sqrt{A}}{d\lambda^2}+\sx^2
=-\frac{1}{2}R_{ij}k^ik^j \label{sqrta} \\
&&\frac{d(A\sx_{ab})}{d\lambda}=AC_{ijkl}L^i_ak^jk^kL^l_b.
\label{asab} \eea

In this work we study light propagation in
space-times with spherical spatial symmetry. In such cases, the
off-diagonal elements $\sx_2$ of the shear tensor can be set consistently
to zero. The reason is that the eigenvalues $\pm \sx$ of the 
shear tensor determine the deformation of a surface along two principal
orthogonal
axes perpendicular to the light direction. The rate of stretching 
in these two directions is given by $\theta+\sx$ and $\theta-\sx$, 
respectively. The off-diagonal elements $\sx_2$ are related to the
orientation of the principal axes with respect to the 
space-like unit vectors $L^i_1, L^i_2$.
In the case of spherical symmetry, one principal
axis lies on the plane determined by the null geodesic and the center of
symmetry and the second perpendicularly to it. It is possible then
to take  $L^i_1, L^i_2$ along the principal directions.
This leads to $\sx_2=0$, which is equivalent to 
$C_{ijkl}L^i_1k^jk^kL^l_2=0$. 
We have checked explicitly that this condition is satisfied for all the
space-times we consider in this work. 
As a result, eq. (\ref{shr}) becomes
\be
\frac{d\sx}{d\lambda}+2\theta \sx=C_{ijkl}L^i_1k^jk^kL^l_1,
\label{shrr} \ee
and eq. (\ref{asab})
\be
\frac{d(A\sx)}{d\lambda}=AC_{ijkl}L^i_1k^jk^kL^l_1.
\label{as} \ee
The basic optical equations for our study are eqs. (\ref{expr}), (\ref{shrr}),
or equivalently eqs. (\ref{sqrta}), (\ref{as}).

\section{Appendix B}
\setcounter{equation}{0}

In the appendix we derive the form of the optical equations 
(\ref{thaa})--(\ref{as}) in specific 
backgrounds.

\subsection{Lemaitre-Tolman-Bondi (LTB) background}

A pseudo-orthonormal basis of vectors along a null geodesic is
\bea
&&k=(k^0,k^1,0,k^3)\label{po1}\\
&&L_1=\left( 0,0,\frac{1}{R},0\right )\label{po2}\\
&&L_2=\left( \frac{hk^0R+k^1b}{k^3R},\frac{k^0+hRk^1b}{k^3bR},0,h
\right) \label{po3}\\
&&w=\left( \frac{k^0+hR(2k^1b+hk^0R)}{2\left(k^3\right)^2R^2},
\frac{k^1b(h^2R^2+1)+2hk^0R}{2\left(k^3\right)^2bR^2},0,
\frac{h^2R^2-1}{2k^3R^2}\right)
\label{po4} \eea
where the function $h(r,t,u)$ in arbitrary. 
For these vectors to be parallely propagated along the light path, the function
$h$ must satisfy the differential equation
\bdm
\frac{dh}{d\lx}=-\frac{1}{R^2b}\left[2bhR
\left(\dot{R}k^0+R'k^1\right)+\dot{R}b^2k^1+k^0R'\right].
\edm
The explicit form of the solution is not necessary for our purposes, as the 
quantities 
\bea
&&R_{ij}k^ik^j=\frac{1}{2M^2}\rho \left(k^0\right)^2\label{ex1}\\
&&C_{ijkl}L_1^ik^jk^kL_1^l=\frac{\left(k^3\right)^2}{4 M^2 R^2}
\left(\rho-\frac{3\calm(r)}{4\pi R^3}\right)\label{ex2}\\
&&C_{ijkl}L_1^ik^jk^kL_2^l=0\label{ex3}
\eea
do not depend on $h$. 
As remarked in the previous appendix, we can set $\sx_2=0$. 
The evolution of the expansion $\theta$ and the shear $\sx=\sx_1$ of a beam
is given by eqs. (\ref{exx1}), (\ref{exx2})
while the beam area evolves according to eq. (\ref{exx3}).

\subsection{Schwarzschild background}

A pseudo-orthonormal basis of vectors along a null geodesic is
\bea
&&k=(k^0,k^1,0,k^3)\label{ps1}\\
&&L_1=\left( 0,0,\frac{1}{r},0\right )\label{ps2}\\
&&L_2=\left( 
\frac{hk^0}{k^3}+\frac{k^1}{k^3 r}
\left(1-\frac{r_s}{r} \right)^{-1},
\frac{hk^1}{k^3}+\frac{k^0}{k^3 r}
\left(1-\frac{r_s}{r} \right),
0,h
\right) \label{ps3}\\
&&w=\frac{1}{2\left(k^3\right)^2r^2}
\Biggl( k^0\left( h^2r^2+1\right) +{2hrk^1}
\left({1-\frac{r_s}{r}}\right)^{-1},
\nn \\
&&~~~~~~~~~~~~~~~~~k^1\left( h^2r^2+1 \right)
+2hrk^0\left(1-\frac{r_s}{r}\right),
0,k^3 \left( h^2r^2-1 \right)
\Biggr),
\label{ps4} \eea
where the function $h(r,t,u)$ in arbitrary. 
For these vectors to be parallely propagated along the light path, the function
$h$ must satisfy the differential equation
\bdm
\frac{dh}{d\lx}+2\frac{hk^1}{r}+\frac{k^0}{r^2}
\left(1-\frac{r_s}{r} \right)=0.
\edm

We have 
\bea
&&R_{ij}k^ik^j=0\label{exz1}\\
&&C_{ijkl}L_1^ik^jk^kL_1^l=-\frac{3\left(k^3\right)^2}{2}
\frac{r_s}{r}\label{exz2}\\
&&C_{ijkl}L_1^ik^jk^kL_2^l=0\label{exz3},
\eea
and we can set $\sx_2=0$. 
The evolution of the expansion $\theta$ and the shear $\sx=\sx_1$ of a beam
is given by eqs. (\ref{exs1}), (\ref{exs2})
while the beam area evolves according to eq. (\ref{exs3}).

\end{document}